\documentclass[preprint,tightenlines,eqsecnum,floats,aps,amsmath,amssymb,nofootinbib,prd,showpacs]{revtex4}

\usepackage{amssymb}
\usepackage{stmaryrd}
\usepackage{amsmath}
\usepackage{amsfonts}
\usepackage{mathrsfs}
\usepackage{CJK}
\usepackage{amsmath,amssymb,amsfonts}
\usepackage{graphicx}
\usepackage{subfigure}
\usepackage{color}

\def\be{\begin{equation}}
\def\ee{\end{equation}}
\def\ba{\begin{eqnarray}}
\def\ea{\end{eqnarray}}
\def\nn{\nonumber}

\begin{document}

\title{Connections between the shadow radius and the quasinormal modes of Kerr-Sen black hole}

\author{Xianglong Wu}
\affiliation{Department of Physics, South China University of Technology, Guangzhou 510641, China}

\author{ Xiangdong Zhang\footnote{Corresponding author. scxdzhang@scut.edu.cn}}
\affiliation{Department of Physics, South China University of Technology, Guangzhou 510641, China}

\date{\today}


\begin{abstract}

The correspondence between the shadow radius and the quasi-normal modes of the Kerr-Sen black hole is studied.  By using the equation of the shadow radius of the Kerr-Sen black hole and the angular separation constant of the quasi-normal modes, the expression of the real part of quasi-normal modes related to the shadow radius is obtained. We found that in the eikonal limit, our formula coincides with the previous result. Moreover, we also calculate the energy emission rate of the Kerr-Sen black hole, it turns out that the peak energy emissivity is inversely correlated with the charge parameter.

\end{abstract}
\maketitle
\section{INTRODUCTION}

In $2019$ the Event Horizon Telescope (EHT) Cooperation released the first black hole shadow image of M87* \cite{L1,L2,L3,L4,L5,L6}, and later in $2021$ the EHT  Cooperation further released its image with polarized light.  In addition to the influence of the observer's angle, the shadow also contains information such as the mass and the rotating parameters of the black holes. Moreover, the black hole shadow can serves as a tool to test the theory of gravity and thus receiving increasingly attention. On the other hand, the first gravitational wave event was observed by The LIGO Scientific Collaboration in 2015 \cite{GW150914}. This is the first time that the gravitational waves has been observed in history, which cast a great influence and promotion on science. The characteristic modes of an exponentially decaying "ringdown" phase of the gravitational waves are described by quasi-normal modes(QNMs), and can be decomposed as $\omega_{QNM}=\omega_{R}-i\omega_{I}$ where the real part $\omega_{R}$ represents the frequency of the wave(oscillation), and the imaginary part $\omega_{I}$ represents the damping.

Though the QNM and black hole shadow seem irrelevant to each other at the first glance. They exist a deep connection. In 2009, by using the WKB approximation method \cite{W1,K1,B1} to calculate of QNMs, V. Cardoso et al.\cite{VCardoso} established the first concrete correspondence between QNMs and black hole shadows for the static spherically symmetric asymptotically flat black hole in the eikonal limit, $\ell \gg 1$ ($\ell$ is the integer angular number). Their results \cite{VCardoso} show the real part of the QNM corresponding to the frequency of the circular null geodesic, while the imaginary part corresponds to the Lyapunov constant that determines the scale of orbital instability. However, although such correspondence formula looks very elegant, it is soon to be found not applicable to some modified gravity theory \cite{RAKonoplya}. Another question is how to extend the correspondence to the rotational black hole case, since most astrophysical real black hole process rotations. More recently, the relations between the QNM and black hole shadow for rotating black hole are proposed by H. Yang et al.\cite{HYang} and K. jusufi \cite{Kimet Jusufi,Jusufi2} respectively. Their results are coincide in the large $\ell$ limit. As examples they study the Kerr, Kerr-Newmann, as well as the five dimensional Myers-Perry black holes \cite{HYang,Kimet Jusufi,Jusufi2,EKN}. However, generalizing this correspondence to a more general case to test its domain of validation is still curial, especially for the black hole in modified gravity theory.

Kerr-Sen (KS) black hole\cite{A. Sen} is an exact  black hole solution in the low-energy effective field theory of the heterotic string theory \cite{A. Sen}, it represents a charged and rotating black hole. Many aspects of KS black holes have been investigated in the past three decades \cite{Larranaga,Scalarclouds,Bernard,SCRCC}. Particularly, the shadow of the KS black hole has been calculated in Ref. \cite{Xavier}, and thus KS black hole serves as an ideal extension to further verify the correspondence between the shadow of a rotational black hole and its QNM. For the strong motivation given in the above, we are going to study the connection between the shadow radius and the QNMs in KS spacetime.

This article is organized as follows. Sec.I is an introduction. Then we give the basic equations of motion for KS black hole in Sec.II. We further calculate the shadow radius by using the unstable photon orbit In Sec.III. In Sec.IV, we analyze the perturbations of the massless scalar field and discuss the angular separation constant. After that, the relation between the QNM and the shadow radius is established in Sec.V and in Sec.VI we give a brief calculation about the energy emission rate of the KS black hole. The conclusion is given in Sec.VI. Throughout this paper, we adopt the geometric units such that $G=c=\hbar=1$.

\section{BASIC EQUATIONS}

\subsection{Basic Equations}
The author in \cite{A. Sen} construct an exact classical black hole solution in the low-energy effective field theory of the heterotic string theory. Note that the low-energy effective action of heterotic string theory in four dimensions can be expressed as
\ba
S=-\int d^{4}x \sqrt{-\widetilde{g}}e^{-\Phi}\left(-R+\frac{1}{12}H_{\mu \nu \rho}H^{\mu \nu \rho}-g^{\mu \nu}\partial_{\mu}\Phi\partial_{\nu}\Phi+\frac{1}{8}F_{\mu \nu}F^{\mu \nu}\right),
\ea
where $\widetilde{g}$ is a determinant of metric tensor $g_{\mu\nu}$, $R$ is the Ricci scalar, $\Phi$ is the dilaton field, $F_{\mu \nu}=\partial_{\mu} A_{\nu}-\partial_{\nu} A_{\mu}$ corresponding to the Maxwell field. Moreover, $H_{\mu\nu\rho}$ is the third-rank field defined as
\ba
H_{\mu\nu\kappa}=\partial_{\kappa}B_{\mu\nu}+\partial_{\mu}B_{\nu\kappa}+\partial_{\nu}B_{\kappa\mu}-\frac{1}{4}\left(A_{\kappa}F_{\mu\nu}+A_{\mu}F_{\nu\kappa}+A_{\nu}F_{\kappa\mu}\right),
\ea
with $B_{\mu\nu}$ being a second-rank antisymmetric tensor field.

Such a theory admits a four-dimensional black hole solution referred to as the Kerr-Sen black hole which is characterized by its mass $M$, charge $Q$ and rotating parameter $a$. In Boyer-Lindquist coordinates $(t,r,\theta,\varphi)$, the metric of Kerr-Sen black hole reads \cite{A. Sen}
\ba
ds^2&=&-\left(1-\frac{2Mr}{\rho^2}\right)dt^2+\rho^2\left(\frac{dr^2}{\Delta_{KS}}+d\theta^2\right)-\frac{4Mra}{\rho^2}sin^2\theta dtd\varphi\\
&&+\left(r(r+2b)+a^2+\frac{2Mra^2sin^2\theta}{\rho^2}\right)sin^2\theta d\varphi^2,
\ea where
\ba
 \Delta_{KS}=r^2-2Mr+2br+a^2,\quad \rho^2=r^2+2br+a^2cos^2\theta,
\ea
 Here the relation between the parameter $b$ and charge $Q$ is $b=Q^{2}/2M$.
The horizon of the KS black hole is determined by $\Delta_{KS}=0$, and therefore is given by $r_{\pm}=M-b\pm \sqrt{(M-b)^{2}-a^2}$.

The geodesic Hamilton-Jacobi equation of KS black hole reads\cite{CBlaga}
\ba
\label{HJE}
\frac{\partial S}{\partial \sigma}=-\frac{1}{2}\frac{\partial S}{\partial x^{\mu}}\frac{\partial S}{\partial x^{\nu}},
\ea
where $S$, $\sigma$ is the principle function and an affine parameter, respectively.
For the null geodesics, the corresponding principle function $S$ reads\cite{CBlaga}
\ba
\label{principleS}
S(t,r,\theta,\varphi)=-Et+S_{r}(r)+S_{\theta}(\theta)+L_{z}\varphi.
\ea Combining Eq.\eqref{principleS} with Eq.\eqref{HJE}, we get two separated  parts of Hamilton-Jacobi equation\cite{CBlaga,ANarang}
\ba
\label{Stheta}
S_{r}(r)=\pm \int\frac{R(r)}{\Delta_{KS}}dr, \quad S_{\theta}(\theta)=\pm \int\sqrt{\Theta(\theta)}d\theta.
\ea
where
\ba
\label{Rr}
R(r)=(aL_z-E(r(r+2b)+a^2))^2-\Delta_{KS}((L_z-aE)^2+\mathcal{D}),
\ea
\ba
\label{theta}
\Theta(\theta)=\mathcal{D}-cos^2\theta\left(\frac{L_{z}^2}{sin^2\theta}-a^2E^2\right).
\ea
The constant $E$ and $L_z$ are the energy and the angular momentum of the photon respectively. And $\mathcal{D}$ is commonly referred to as the Carter separation constant \cite{BCarter,CBlaga}.

Considering the Hamilton-Jacobi equation \eqref{HJE}, the equations of motion of particles in the KS spacetime are determined by the following four first-order linear differential equations
\ba
\rho^2 \dot t&=&\frac{E(r(r+2b)+a^2)^2-2MraL_z}{\Delta_{KS}}-a^2E^2\sin^2\theta,\\
\rho^2\dot\varphi&=&-aE+\frac{L_z}{\sin^2\theta}+\frac{a}{\Delta_{KS}}(r(r+2b)+a^2E-aL_z),
\ea
\ba
\rho^2 \dot\theta&=&\pm \sqrt{\Theta(\theta)},
\ea
\ba
\rho^2 \dot r&=&\pm \sqrt{R(r)},
\ea

\section{Shadow radius of  the Kerr-Sen black hole}
The size and shape of the shadow of a black hole are determined by the unstable circular photon orbit. For the observer at infinity, the observed shape is affected by the inclination angle $\theta_0$ of the observer.  When viewed from the equatorial plane($\theta_0=\pi/2$), we follow the definition in Ref. \cite{XHFeng,MZhang}, for the typical shadow radius can be expressed as
\ba
\label{radium rs}
\overline{R}_{s}=\frac{1}{2}(x(r^{+}_{0})-x(r^{-}_{0})),
\ea
here $x(r^{\pm}_{0})$ denote the unstable photon orbits \cite{XHFeng,Kimet Jusufi}.

Consider the circular unstable photon orbit in the equatorial plane($\theta=\pi/2$),  the appropriate Lagrangian is
\ba
\mathcal{L}=\frac{1}{2}g_{\mu\nu}\dot{x^{\mu}}\dot{x^{\nu}}.
\ea

\begin{figure}[!htb]
	\includegraphics [width=0.5\textwidth]{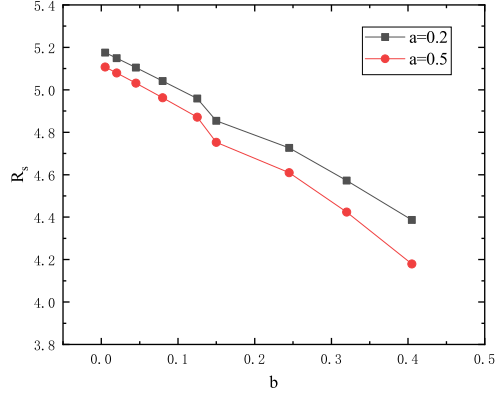}
	\caption{The shadow radius of KS black hole for different values of $b$ when the rotating parameter chosen as a=0.2 or a=0.5. }
	\label{shadowradiusa0205}
\end{figure}

Since the Kerr-Sen spacetime is stationary and axially symmetric, from the conserved quantities of the test particle, we can conclude that
 \ba
p_{t}=\frac{\partial \mathcal{L}}{\partial \dot t}&=&g_{tt}\dot t+g_{t\varphi}\dot \varphi=E,  \\
p_{\varphi}=\frac{\partial \mathcal{L}}{\partial \dot \varphi}&=&g_{t\varphi}\dot t+g_{\varphi \varphi}\dot \varphi=-L_z,\\
p_{r}=\frac{\partial \mathcal{L}}{\partial \dot r}&=& g_{rr}\dot r.
\ea
For the null geodesics in the equatorial plane, the corresponding Hamiltonian is
\ba
\label{geodesic delta}
H=p_{t}\dot t+p_{\varphi}\dot \varphi+p_{r}\dot r-\mathcal{L}=0.
\ea
Here we will introduce $V_{eff}$ as the effective potential of the photon\cite{VCardoso} which is defined as
\ba
V_{eff}=\dot r^2.
\ea
Then combine the above definition with Eq. \eqref{geodesic delta} we obtain
\ba
V_{eff}=-\frac{g_{tt}\dot t^2+2g_{t\varphi}\dot t\dot \varphi+g_{\varphi\varphi}\dot\varphi^2}{g_{rr}}.
\ea For a circular photon orbit it is required that \cite{HPress}
\ba
 \label{partial effective petential}
V_{eff}=0, \quad V^{'}_{eff}=0.
\ea
By simplifying Eqs.\eqref{partial effective petential} we have the following two equations:
\ba
 \label{Veff 1}
\left(a^2 (2 b+2 M+r)+r (2 b+r)^2\right)E^2-4aL_{z}ME-L_{z}^2 (2 b-2 M+r)=0,
\ea
and
\ba
 \label{Veff 2}
-a^2 E^2 M+2 a E L_{z} M+4 b^3 E^2+8 b^2 E^2 r+5 b E^2 r^2+E^2 r^3-L_{z}^2 M=0.
\ea
 Note that the black hole shadow radius $R_s$ can be expressed as $R_{s}=L_{z}/E$ \cite{Kimet Jusufi}. Then Eq. \eqref{Veff 2} equivalent to
\ba
(R_{s}-a)^{2}M-(8b^{2}r+5br^{2}+r^{3}+4b^3)=0,
\ea
which implies that
  \ba
  \label{Rs pm}
 R^{\pm}_{s}=a\pm \sqrt{\frac{4b^3+8b^{2}r^{\pm}_{0}+5b(r^{\pm}_{0})^{2}+(r^{\pm}_{0})^{3}}{M}}.
 \ea
 Here $r^{\pm}_{0}$ are the solutions of Eq.\eqref{Veff 1} by substituting Eq.\eqref{Rs pm} into Eq.\eqref{Veff 1}. In such a situation, the definition in Eq.\eqref{radium rs} equals to
 \ba
\overline{R}_{s}=\frac{1}{2}(R^{+}_{s}\vert_{r^{+}_0}-R^{-}_{s}\vert_{r^{-}_0}).
\ea
Then the typical shadow radius for KS black hole reads
\ba
\overline{R}_{s}=\frac{1}{2} \left(\sqrt{\frac{4b^3+8b^{2}r^{+}_{0}+5b(r^{+}_{0})^{2}+(r^{+}_{0})^{3}}{M}}+\sqrt{\frac{4b^3+8b^{2}r^{-}_{0}+5b(r^{-}_{0})^{2}+(r^{-}_{0})^{3}}{M}}\right).
\ea
It is a function of the black hole mass $M$, the rotating and the charge parameter $a$, $b$ respectively. In this paper, we have set the black hole mass $M=1$. From Fig. \ref{shadowradiusa0205} it can be concluded that the shadow radius decrease with the increase of the charge parameter $b$. And for the same value of $b$, the larger of the rotational parameter $a$ corresponds to the smaller radius.

\section{QNMs of the Kerr-Sen black hole}
\subsection{Perturbation of the scalar field}
Considering a massless scalar field $\Phi$ in the KS spacetime, it satisfied the Klein-Gorden equation
\ba
\frac{1}{\sqrt{-g}}\partial_{\alpha}(g^{\alpha\beta}\sqrt{-g}\partial_{\beta}\Phi)=0.
\ea Teukolsky's work \cite{Teukolsky} shows that all the scalar fields which satisfied $\nabla^{2}\Phi=0$ are separable in Boyer-Lindquist coordinates. Such that for the scalar field $\Phi(t,r,\theta,\varphi)$ in KS spacetime, we can decompose it as
  \ba
  \label{phasephi1}
  \Phi=e^{-i\omega t}e^{im\varphi}R(r)S(\theta),
  \ea
It is easy to see that in the eikonal limit, $R(r)$ and $S(\theta)$ satisfied the following equations:
\ba
\label{function S(theta)}
\frac{1}{\sin\theta} \frac{d}{d\theta}\left (\sin\theta\frac{dS(\theta)}{d\theta}\right)+\left(a^{2}\omega^{2}cos^{2}\theta-\frac{m^2}{\sin^{2}\theta}+A_{\ell m}\right )S(\theta)&=&0,\\
\label{function R(r)}
\frac{d}{dr}\left(\Delta_{KS}\frac{dR(r)}{dr}\right)+\left(\frac{(r(r+2b)+a^2)^2}{\Delta_{KS}}+2am\omega-A_{\ell m}\right)R(r)&=&0,
\ea
where $m$ and $A_{\ell m}$ are the azimuthal quantum number and the angle eigenvalue, respectively. $A_{\ell m}$ is a function of $\omega$. When the parameter $a$ and $b$ take to zero(the Schwarzschild limit), the angle eigenvalue takes a simple form as $A_{\ell m}=\ell(\ell+1)$. In general, the expression of $A_{\ell m}$ is quite complicated and we can separate $A_{\ell m}$ into real and imaginary part \cite{yhuang}
\ba
A_{\ell m}=A^{R}_{\ell m}-iA^{I}_{\ell m}.
\ea
For Eq. \eqref{function S(theta)} it should satisfy the Bohr-Sommerfeld quantization condition \cite{yhuang}
\ba
\label{BSommer}
\int^{\theta_+}_{\theta_-}{\sqrt{a^{2}\omega^{2}_{R}cos^{2}\theta-\frac{m^2}{sin^{2}\theta}+A^{R}_{\ell m}}}d\theta=(L-\mid{m}\mid)\pi.
\ea
Here we have set $L=\ell+1/2$ \cite{yhuang}. $\theta_{\pm}$ are the turning point and the zero point of the potential respectively. For the angular equation Eq.\eqref{function S(theta)}, motivated by the WKB analysis, defining $dx=d\theta/sin\theta$, $x=log(tan\frac{\theta}{2})$, the Eq. \eqref{function S(theta)} can be rewritten as
\ba
\label{functionWKB}
\frac{d^{2}S(\theta)}{dx^2}+(a^{2}\omega^{2}_{R} cos^{2}\theta sin^{2}\theta-m^2+A^{R}_{\ell m}sin^{2}\theta)S(\theta)=0.
\ea
Eq.\eqref{function S(theta)} has two regular singular points, $cos\theta=+1$ and $cos\theta=-1$. The boundary condition for Eq.\eqref{function S(theta)} is that $S_{\theta}$ be finite at the singular points. Similar to E. W. Leaver\cite{Leaver}, we can write a solution to Eq.\eqref{function S(theta)} as
\ba
\label{three term}
S_{\theta}=e^{a\omega cos\theta}(1+cos\theta)^{\frac{|m|}{2}}(1-cos\theta)^{\frac{|m|}{2}}\sum^{\infty}_{n=0}a_{n}(1+cos\theta)^{n}.
\ea By substituting Eq. \eqref{three term} into Eq. \eqref{function S(theta)}, we can get a three-term recurrence relation:
\ba
&&\alpha_{0}a_1+\beta_{0}a_0=0.\\
&&\alpha_{n}a_{n+1}+\beta_{n}a_{n}+\gamma_{n}a_{n-1}=0.  \quad   n=1,2...
\ea

For a given value of $a$ and $m$, $\omega_{R}$ and $A_{\ell m}$ can be found by solving the continued fraction equation. We find that the form and recurrence coefficient of this recurrence relations are the same as that in the case of the Kerr black hole scalar field\cite{Leaver}, for they have the same form equation and boundary condition of the massless scalar field's angular separation function. Therefore, for the angular separation constant $A_{\ell m}$ in the KS black hole scalar field equation can be obtained by following the same analysis and calculation in Ref \cite{Berti,yhuang}. Particularly, in the eikonal limit with $\ell \gg 1$, we can take $a\omega _R$ as a small value. Then the separation constant $A^{R}_{\ell m}$ can be expanded as a Taylor series \cite{Leaver}:
\ba
\label{Alm}
A^{R}_{\ell m}=\sum^{\infty}_{p=0} f_{p}(a\omega_R)^{p}\approx f_{0}+f_{2}(a\omega_R)^{2}+O(a\omega_R)^{4}\approx L^{2}+\frac{1}{2}(\frac{m^2}{L^2}-1)a^{2}\omega^{2}_{R}.
\ea
\subsection{Connections between the QNMs and shadow radius}

By identifying the massless scalar field $\Phi$ with the leading order of the principle function \eqref{principleS}, we can rewrite it as
\ba
\label{phasephi2}
\Phi=e^{iS}=e^{-iE t}e^{iL_z \phi}e^{iS_{\theta}}e^{iS_r}.
\ea
After comparing Eq.\eqref{phasephi1} with Eq.\eqref{phasephi2}, it is not difficult to conclude that
\ba
\label{ELZ}
E=\omega_R, \quad L_{z}=m.
\ea

Moreover, when considering equations  \eqref{Stheta}, \eqref{function S(theta)} and \eqref{functionWKB}, and use the WKB method used in Ref. \cite{yhuang}, we can further make the identification that

\ba
\label{mathcalD}
\mathcal{D}=A^{R}_{\ell m}-m^2.
\ea
For typical QNMs, it can be expressed as \cite{yhuang}
\ba
\omega=(\ell+\frac{1}{2})\Omega_{R}(\mu)-i(n+\frac{1}{2})\Omega_{I}(\mu),
\ea
with $ \mu \equiv m/(\ell+1/2)$.

For a rotating black hole, we introduce a new angle $\Delta\varphi_{prec}$ \cite{BMashhoon,yhuang}, which represents the Lense-Thiring-precession frequency of the orbit that arises because of the rotation of the black hole \cite{yhuang}. If we define $T_{\theta}$ as a period of motion in the $\theta$ direction, then the corresponding precession frequencies $\Omega_{prec}$ is
\ba
\label{oprec}
\Omega_{prec}=\frac{\Delta\varphi_{prec}}{T_{\theta}}.
\ea
Such that in the rotating black hole the real part of the frequency can be written as\cite{HYang}
\ba
\Omega_{R}=\Omega_{\theta}(\mu)+\mu\Omega_{prec}(\mu),
\ea
with $\Omega_{\theta}=2\pi/T_{\theta}$.

 Considering a complete cycle of the photon orbit in the $\theta$ direction,
\ba
\label{principlrS}
\delta S=L_{z}\Delta \varphi-ET_{\theta}+\delta S_{\theta}=0,
\ea
where  $\Delta\varphi$ is the azimuth changed after completing a cycle in the direction, it relates to the $\Delta \varphi_{prec}$ by
\ba
\Delta \varphi=\Delta \varphi_{prec}+2\pi sgn(L_z),
\ea
where sgn(.) evaluates the sign of the argument.
For $\delta S_{\theta}$, consider it with Eq.\eqref{BSommer}, we can get the equation that
\ba
\label{T2}
\delta S_{\theta}=2\int^{\theta_+}_{\theta_-}{\sqrt{\Theta}d\theta}=2\int^{\theta_+}_{\theta_-}{\sqrt{\mathcal{D}-cos^2\theta\left(\frac{L^2}{sin^2\theta}-a^2E^2\right)}d\theta}=2\pi(L-L_z).
\ea
Combining Eqs.\eqref{ELZ},\eqref{oprec}-\eqref{T2}, we found that
\ba
\label{L/Eandomega}
\frac{L}{E}=\frac{1}{\Omega_\theta+\mu\Omega_{prec}}=\frac{1}{\Omega_R}
\ea
By substituting $\mathcal{D}=A^{R}_{\ell m}-m^2$ into Eq.\eqref{Alm}, in large $\ell$ case, there is a equation that

\ba
\label{1/2}
\frac{\sqrt{\mathcal{D}+L^{2}_{z}}}{E}\approx R_{s}.
\ea Moreover, combining Eqs. \eqref{Alm}, \eqref{mathcalD} and \eqref{1/2}, we can get that
\ba
\label{L/E and RS}
\frac{L^2}{E^2}&=&\frac{\mathcal{D}+L^{2}_{z}}{E^2}+\frac{a^2}{2}\left(1-\frac{m^2}{L^2}\right)\nn\\
&\approx&R_{s}^{2}+\frac{a^2}{2}\left(1-\frac{m^2}{L^2}\right),
\ea
Collecting all the ingredients, for the shadow radius $R^{+}_s$ and $R^{-}_s$, the connection between the real part of the QNMs and the shadow radium is \cite{Kimet Jusufi} \ba
\omega_{R}=\frac{1}{2}\left(\omega_{R^{+}}-\omega_{R^{-}}\right)
\ea with $\omega_{R^{\pm}}$ being
\ba
\label{omegaR}
\omega_{R^{\pm}}=\pm \frac{\ell+\frac{1}{2}}{\sqrt{(R^{\pm}_{s})^{2}+\frac{a^2}{2}(1-\mu^2)}}.
\ea
 When taking the limit that $\ell \gg 1$, and $m=\pm \ell$ which means $\mu \to 1$. Then the above equation can be reduced to $\omega_{R^{\pm}}\approx\frac{\ell}{R^{\pm}_{s}}$, which agrees with the statement in Ref.\cite{Kimet Jusufi}.

We can get the real part of QNMs of KS black by the use of the correspondence equation \eqref{omegaR}. Table \eqref{table1} shows the QNMs of the black hole obtained by applying geometrical optics approximation to calculate the shadow of the black hole under different parameters $b$. With the increase of parameter $b$, the corresponding shadow radius of the black hole decreases gradually and the value of QNMs increases. Comparing our results with those of Kerr-Newman(KN) black hole in Ref. \cite{Kimet Jusufi}, we found that the black hole shadow radius of KS black hole is always slightly larger than that of KN black hole under the same charge and rotation parameter. This is consistent with the conclusion in Ref. \cite{Xavier}.
  \begin{table}
	\renewcommand{\arraystretch}{1.3}
	\centering
\begin{tabular}{lccr}
\hline
$b$&$\omega^{+}_{R}$&$\omega^{-}_{R}$&$\overline{R}_s$\\
\cline{1-4}
0.005&16.81283709&-22.97778200&5.17568286\\
\cline{1-4}
0.02&16.88189993&-23.12877614&5.14917896\\
\cline{1-4}
0.045&16.99831595&-23.38742770&5.10476553\\
\cline{1-4}
0.08&17.16644714&-23.76637234&5.04155388\\
\cline{1-4}
0.125&17.39119052&-24.28492470&4.95857888\\
\cline{1-4}
0.18&17.68007277&-24.97320521&4.85433954\\
\cline{1-4}
0.245&17.04378505&-25.87821277&4.72668035\\
\cline{1-4}
0.32&18.49758147&-27.07611124&4.57245092\\
\cline{1-4}
0.405&19.06366779&-28.69898429&4.38683720\\
\hline
\end{tabular}
 \caption{shadow radius of KS black hole for different charge parameter $b$ when $a=0.2$. }
 \label{table1}
  \end{table}
\section{Energy emission rate}
Next, let's focus on the emission of particles in KS spacetime. Generally speaking, the numerical oscillation range of the absorption cross-section of a spherically symmetric black hole is the limit constant $\sigma_{lim}$\cite{BSMashhoon}. From the point of view of the observer located at the infinity, the absorption section of the black hole corresponds to the shadow of the black hole. For a spinning black hole, its shadow can also be roughly seen as a circle \cite{EMRapprox}. So the limit constant can be approximately expressed as\cite{AFolacci}
\ba
\sigma_{lim}\approx\pi \overline{R}^{2}_s.
\ea
Therefore, the energy emission rate of KS black hole at high energy reads
\ba
\frac{d^{2}E}{d\varpi dt}=\frac{2\pi^{3}\overline{R}^{2}_s}{e^{\varpi/T}-1}\varpi^{3},
\ea
where $\varpi$ is the emission frequency and $T$ is the Hawking temperature of the KS black hole which reads\cite{Larranaga}
\ba
T=\frac{\sqrt{(M-b)^{2}-a^2}}{4\pi M(M-b+\sqrt{(M-b)^{2}-a^2})}.
\ea
\begin{figure}[!htb]
	\includegraphics [width=0.6\textwidth]{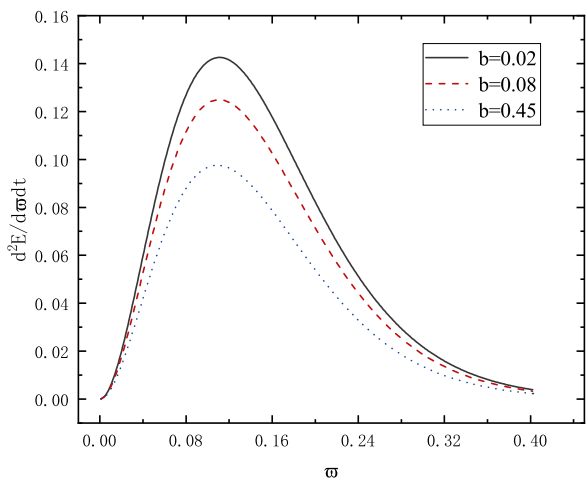}
	\caption{The energy emission rate of KS black hole for different charged parameter $b$ with a fixed rotating parameter $a=0.2$. The three lines from top to bottom corresponding to $b=0.02, 0.08$ and $0.45$, and the peak decreasing step by step. }
	\label{emergy emission rate b=0.02}
\label{figEMR02}
\end{figure}

\begin{figure}[!htb]
	\includegraphics [width=0.6\textwidth]{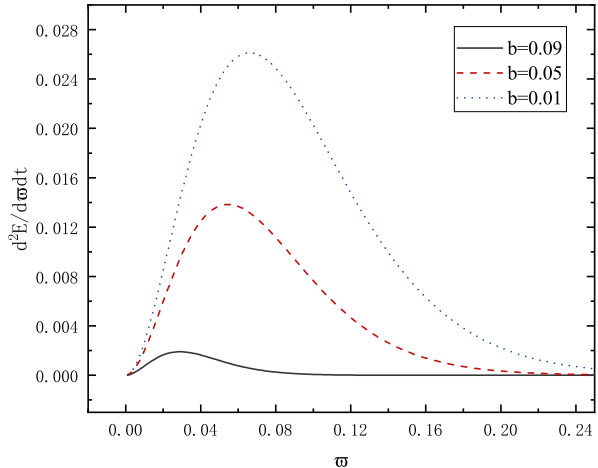}
	\caption{The energy emission rate of KS black hole for different charged parameter $b$ by fixing the rotating parameter $a=0.9$. The three line from the top to bottom corresponding to $b=0.01, 0.05$ and $0.09$. }
    \label{fig09line15}
\end{figure}
From Fig. \ref{figEMR02} and Fig. \ref{fig09line15}, we found that there has a peak in the energy emission rate of the KS black hole. The peak value decreases with the increases of $b$.

\section{CONCLUSIONS}

In this paper, we study the connections between the shadow radius and the real part of QNMs of KS black hole in the eikonal limit, and by using this relationship, we calculate the corresponding real part of QNMs of KS black hole through shadow radius. We first calculate the typical shadow radius by the unstable photon orbit for KS spacetime and found that the shadow radius decreases with the increase of the parameters $a$ and $b$ of KS black holes. Then we discuss the perturbation of the massless scalar field in the KS background. The corresponding field equation turns out to be separable. Compare with the Kerr black hole case, despite the radial equation for $R(r)$ appearing quite different,  the separation function for the $\theta$ direction keeps the same as the case of Kerr black hole due to the same axial symmetry.

 Through the comprehensive analysis of the perturbation of the massless scalar field and the principle Hamilton-Jacobi function, we get the correspondence relation between QNM and shadow radius. Through this correspondence, we calculate the QNMs of KS black hole. In the eikonal limit, our result reduces to the formula $\omega_{R^{\pm}}\approx\frac{\ell}{R^{\pm}_{s}}$ obtained by the approximation method in Ref. \cite{Kimet Jusufi}. This result confirms that the real part of the QNMs corresponding to the unstable circular photon orbit is still valid for KS black holes. Moreover, in the last part of this paper, the energy emission rate of the KS black hole has also been studied, we found that the peak of emission rate decreases when the charge parameter $b$ increases for a given rotation parameter $a$.

\begin{acknowledgements}
This work is supported by NSFC with Grants No. 11775082.

\end{acknowledgements}

\bibliographystyle{unsrt}

\end{document}